\documentstyle[preprint,pra,aps]{revtex}

\tightenlines

\begin{document}
\draft
%
%--------------------------------------------------------------------
%
\title{Short distance relativistic atom-atom forces\thanks{Contribution
for third issue of the Bulletin of the  Asia Pacific Center for
Theoretical Physics, Seoul, Korea.}}

\author{J.~F. Babb}
\address{
Institute for Theoretical Atomic and Molecular Physics,\\
Harvard-Smithsonian Center for Astrophysics,\\
60 Garden Street, Cambridge, Massachusetts 02138
}
\date{March 1, 1999}
\maketitle
%--------------------------------------------------------------------
%\begin{abstract}
%For the interactions between two atoms or between
%an ion and an electron the connections between the Casimir-Polder-type
%potentials and the relativistic potentials arising from the Breit-Pauli
%Hamiltonian are surveyed.
%\end{abstract}
%
%--------------------------------------------------------------------
%

About fifty years ago two important papers appeared describing novel
interactions. One, by Casimir, discussed the case of two interacting
walls~\cite{Cas48} and the other, by Casimir and Polder, considered
the interactions between an atom and a wall and between two
atoms~\cite{CasPol48}.
The history and some of the many interesting aspects of these
interactions, their derivations, and their importance in field theory
and atomic and molecular physics are summarized
elsewhere~\cite{Mil93,Spr96,MosTru97}.  Indeed, recent experiments give
strong evidence of the reality of both the
atom-wall~\cite{SukBosCho93} and the wall-wall
interactions~\cite{Lam97} predicted in those two papers.

Here I will focus on the connection between the QED result of Casimir
and Polder and other results for relativistic atom-atom interactions at
short distances of the order of, say, $20~a_0$.  The interaction
between an electron and an ion will also be considered. For typical
atomic systems these relativistic effects are very small corrections
to the non-relativistic potentials arising from the van der Waals and
Coulomb interactions for, respectively, the atom-atom interaction and
the electron-ion interaction.

Casimir and Polder used QED and old-fashioned perturbation theory
and their result was subsequently duplicated by other authors with
different methods, cf.~\cite{Mil93}.  
One way to write their result for the
interaction potential $V(R)$ between two spherically symmetric atoms
separated by a distance $R$ is as a one-dimensional integral
\begin{equation}
	V(R) = -\frac{1}{\pi R^6} 
       \int_0^\infty 
       d\omega \exp (-2\alpha\omega R)
       [\alpha_d(i\omega)]^2 P(\omega\alpha R),
	\label{CP}
\end{equation}
with $P(x) = x^{4}+2x^{3}+5x^{2}+6x+3$ and $\alpha$ the fine structure
constant.
Atomic units $\hbar= m= e=1$ are used throughout 
and in these units $c=1/\alpha$.
The function $\alpha_d (i\omega)$ is the dynamic electric
dipole polarizability at imaginary frequency,
\begin{equation}
	\alpha_d(i\omega)  = \sum_u f_{u}/[(E_u-E_0)^2+\omega^2],
	\label{polariz}
\end{equation}
where $f_u$ is the oscillator strength of state $u$ and $E_u-E_0$ is
the transition frequency between the states $u$ and $0$, 
with $0$ denoting the ground state of the atom, and the summation
in (\ref{polariz}) includes an integration over continuum states. The
function $\alpha_d(i\omega)$ is a smooth function of $\omega$ with no
singularities.
The limit of $V(R)$ for 
asymptotically large separations  of the atoms
can be obtained from the  Casimir-Polder integral (\ref{CP}) yielding
\begin{equation}
        V(R) \rightarrow -\frac{23}{4\pi}\frac{[\alpha_d (0)]^2}{\alpha R^7}, 
       \qquad R\rightarrow\infty .
       \label{CP-asymp}
\end{equation}

What about the limit for small $R$?  The result is
\begin{equation}
	V(R) \rightarrow -\frac{3}{\pi R^6} 
      \int_0^\infty d\omega \; [\alpha_d(i\omega)]^2, \qquad R\ll 137,
	\label{aa-small}
\end{equation}
and upon integration (\ref{aa-small}) yields 
\begin{equation}
	V(R) \rightarrow -C_6/R^6, 
	\label{aa-small-closed}
\end{equation}
where $C_6$ is the van der Waals constant
expressed as a double sum
over oscillator strengths and the correct form of the
atom-atom interaction at short distances (say $20~a_0$) is
reproduced. The van der Waals constant is of vast importance for all
sorts of molecular 
spectroscopic and atomic collision problems, of course.
The result for two H atoms is $C_6= 6.499\,026\,705...$ and
for studies of atomic collisions at ultracold temperatures $C_6$ plays a
crucial role in characterizing the
interactions~\cite{WeiBagZil99}.  So it is nice to see QED connect
nicely with non-relativistic molecular quantum mechanics.  What is the
next correction?

If one more term is retained in the small $R$ expansion then~\cite{MeaHir66}
\begin{equation}
	V(R) =
           -\frac{C_6}{R^{6}} +\alpha^{2}\frac{W_{4}}{R^{4}} 
       + {\cal O} (\alpha^3/R^3)
	\label{aa-rel}
\end{equation}
where
\begin{equation}
	W_{4}= \frac{1}{\pi}\int_{0}^{\infty} d\omega \;
        \omega^2 [\alpha_d(i\omega)]^2 .
	\label{w4}
\end{equation}
By integrating (\ref{w4}), the coefficient $W_4$ can be expressed as a double
sum over oscillator strengths and it was evaluated
for a small number of diatomic systems using various approximations,
both semi-empirical~\cite{MarMea78} and
computational~\cite{YanDalBab97,YanBab98}. The result for two H atoms
is $W_{4}=0.462\,807$.  The derivations above assume that the two atoms are
well-separated and accordingly do not include considerations involving
electron exchange.

How do the results above connect with results from the Breit-Pauli
approximation to the Dirac equation?  The van der Waals potential was
shown above to be the short range limit of the QED result; yet it is also
the {\em long}-range limit of the {\em molecular\/} interaction
potential. The full power of quantum-chemical methods (recognized in
the 1998 Nobel Prize in Chemistry) enables, at least in principle,
calculation of the molecular potential by solution of the
nonrelativistic Schr\"odinger equation.  Relativistic effects are
treated using perturbation theory on the terms in the Breit-Pauli
Hamiltonian (or for molecules containing high-$Z$ atoms by solution of the
Dirac equation.)  The connection to (\ref{aa-rel}) was given by Power
and Zienau~\cite{PowZie57,MeaHir66} who showed using
perturbation theory that the matrix element
of the orbit-orbit interaction $H_{\rm oo}$ reproduces the second
term in 
(\ref{aa-rel}) as $R$ increases,
\begin{equation}
	\langle \tilde{0} | H_{\rm oo} | \tilde{0} \rangle \rightarrow 
        \alpha^{2} \frac{W_{4}}{R^{4}}, \qquad R \sim R_0,
	\label{oo}
\end{equation}
where $R_0$ is of order, say, $10$ to $20~a_0$
and $ | \tilde{0} \rangle$ is the molecular ground electronic
state wave function.
Therefore there is a smooth connection between the relativistic and
Casimir-Polder results.

This relativistic $R^{-4}$ term might be studied by
incorporating it into theoretical
calculations of collisions of ultra-cold atoms, particularly for H-H,
H-Li, and Li-Li where high precision determinations of the molecular
potentials are possible. There are of course additional subtle effects
to be accounted for such as
deviations from the Born-Oppenheimer
approximation through  isotope effects and 
nonadiabatic terms (nonlocal terms arising from the action of the
nuclear kinetic energy operator on the electronic wave function) and additional
relativistic terms like the $p^{4}$ and Darwin terms, for example, but
these are unrelated to the Casimir-Polder result.

%Helium dimer?
%Penfield?
%Hydrogen molecule?

The Casimir-Polder-type interaction between an electron and an ion is closely
related to that of the atom-atom interaction (\ref{CP-asymp}). Kelsey
and Spruch~\cite{KelSpr78} exhibited the result for
asymptotic separations,
\begin{equation}
U (R) \rightarrow  \frac{11}{4\pi}
      \frac{\alpha\alpha_d (0)}{R^5},
\qquad R\rightarrow \infty,
\label{KS}
\end{equation}
where $R$ now denotes the electron-ion distance.  They obtained (\ref{KS})
using QED and old-fashioned perturbation theory and they considered
the possibility of measurement of this potential through spectroscopy
of the Rydberg states of atoms.  Later, the integral form 
of $U(R)$, analogous to
(\ref{CP}), was obtained~\cite{AuFeiSuc84,BabSpr87} yielding
an expression not 
particularly more complicated than (\ref{CP}) and which can be obtained
essentially by replacing the polarizability 
$\alpha_d (i\omega)$ of one of the atoms by the
quantity $1/\omega^2$, which is an excellent
approximation to the polarizability of the weakly bound
electron~\cite{SprKel78}. (Some care is required, however,
due to the additional
Coulomb interaction present for
the ion-electron case, see \cite{BabSpr87} for details.)  The
limit of the electron-ion ``Casimir-Polder'' potential for small $R$
for an electron interacting
with an ion (of net charge $Z-1$) is~\cite{BabSpr88}
\begin{equation}
U(R) = \frac{\alpha^2}{Z^2} \frac{1}{R^4} + {\cal O} (\alpha^3/R^3),
\qquad R \ll 137/Z^2.
\label{ion-small}
\end{equation}
Similarly to the atom-atom case, the relativistic $R^{-4}$ term in the
ion-electron interaction was derived alternatively using $H_{\rm oo}$
with perturbation theory on the non-relativistic wave function of the
Rydberg atom~\cite{Hes92} providing a connection to the QED result
(\ref{ion-small}).  This term is a small correction to the much larger
Coulomb interaction between the two charged particles, but
nevertheless, through much theoretical work by Drachman, Drake, and
others~\cite{LRF}, there is definitive evidence for the
first term of (\ref{ion-small}) from a long series of careful
measurements of energies of Rydberg states of the helium atom by
Lundeen and collaborators~\cite{LRF,ClaHesLun95}.  At present the
asymptotic part of $U(R)$, (\ref{KS}), has not been measured and
Hessels and collaborators~\cite{StoRotHes95} conclude from their
measurements that there is, in fact, no experimental evidence for
deviations from (\ref{ion-small}).  Additional experiments are in
progress~\cite{SteBirLun98} and it will be interesting to see if the
ion-electron Casimir effect will be verified.  From a theoretical
point of view there are interesting connections at short~\cite{Ara57}
and long distance between the order ${\cal O}(\alpha^{3}/R^3)$ QED
corrections in (\ref{aa-rel}) and (\ref{ion-small}).

%\acknowledgements
This work was supported in part by the National Science Foundation
through a grant for the Institute for Theoretical Atomic and Molecular
Physics at the Smithsonian Astrophysical Observatory and Harvard
University.

%--------------------------------------------------------------------

%--------------------------------------------------------------------
\end{document}